# Looking back at Facebook content and the positive impact upon wellbeing: Exploring reminiscing as a tool for self soothing.

Alice Good[1], Arunasalam Sambhanthan, Vahid Panjganj

[1] University of Portsmouth, PO1 3AE, UK
alice.good@port.ac.uk

**Abstract.** The premise of this paper is to explore the potential of reminiscing in facilitating self soothing. The research presented looks at people's activities on Facebook and whether these particular activities impact upon their perceived sense of wellbeing, furthermore, whether specific Facebook activities enable a self –soothing effect when feeling low in mood. A survey was distributed amongst Facebook users. The results from the study appear to indicate that in comparison to other Facebook activities, looking back upon photos and wall posts in particular**,** could have a positive impact upon wellbeing. Additionally, the results indicate that people who have mental health problems, experience a more positive impact upon their wellbeing when looking at photos and wall posts, than those who did not have a history of mental health issues. The results from the research presented here contribute towards the viability of developing a mobile application to facilitate positive reminiscing.

## 1 Introduction

The use of technology in promoting 'well being' has enormous potential, as has been seen through the varied applications developed and utilized over recent years. These include social networking sites, discussion forums, virtual environments such as Second Life and more recently phone apps. Indeed, since the advent of Web 2.0 technologies in 2004, there has been an increase in the use of social networking sites and other applications that enable online communities, which facilitate support for people with mental health problems, or people who simply require occasional emotional support. Social networking sites and discussion forums, are reported to have decreased the sense of 'feeling alone' for people with mental health problems (Neal & McKenzie, 2010; Patti *et al.*, 2007). Facebook usage in particular, is reported to increase a sense of wellbeing amongst its users. The aspects of wellbeing relate to the fact that users were more easily able to form relationships, provide companionship and emotional support, and feel more positive after reading status updates (Hampton et al, 2012; Mauri *et al*., 2011; Burke et al, 2010; Toma, 2010).

The research presented here looks more closely at the extent to which Facebook can impact or improve upon wellbeing. The focus of this paper is to then look specifically at users' interactions on Facebook, and evaluate whether specific interactions, namely looking back upon wall posts and photos, has any impact upon positive wellbeing. The theory behind the research is linked to positive reminiscing, and how this process can impact upon emotional wellbeing.

The results from the study will inform the viability of a mobile application which would enable users to store 'favourite' photos and notes. The rationale being, that facilitating the means for people to easily access their 'favourite things', could potentially increase a sense of wellbeing, particularly among those people who have some history of mental health issues. The research is not claiming that such a tool could cure any mental health issues, but rather that it could serve as a tool to promote self soothing in times of low mood. This research should be considered more as an exploratory study, given the small number of participants, and therefore the results should be viewed as indicative and also as a prelude to further research.

## 2  Social Networking and Wellbeing

The Internet has changed peoples" lives in many aspects, including social and health (Van de belt *et al.*, 2010) and wellbeing. Social Network Sites have been described as a consequence of these changes, which introduced a new way of communication among people (Ross et al., 2009; Cheung & Lee, 2010). Social Network Sites (SNS) are indisputably popular. Facebook was listed as the most visited website, with 800,000,000 unique visitors in a "double click ad planner" report by Google (2013), having reached a billion users in September 2012 (Business Week, 2012). Online Communities are the main active areas in which applications such as Facebook can impact upon a positive sense of wellbeing.

Many people prefer to utilize the wide spectrum of ICT applications to facilitate support, as opposed to face-to face support. For example, face-to-face support groups are often difficult to schedule and are limited in manner of time and location. Moreover, many people with psychiatric problems can experience difficulties accessing these support groups (Taylor & Luce, 2003). In contrast, online support groups and forums are much more accessible, particularly given the growth in smart phones providing ubiquitous access. Studies by Van Uden-Kraan et al (2008) highlight the fact that users within these support groups not only receive support, but are also able to advise and share their own experience, which could potentially provide psychological well-being for the patient. Shepsis (2010) reports interactions made by participants, as more frequent and with higher level of candor comparing to the traditional (offline) methods. These findings are supported by earlier studies, which suggest online communities are a potential option for emotional support (Van Uden-Kraan et al, 2008) and as a consequence, can positively enhance wellbeing

The behavior and type of activities within these SNS provide a new social experience, which can be deeper than traditional ways of socializing, as well as impacting positively upon wellbeing. Examples of this firstly include the way these SNSs facilitate establishing and maintaining pre-existing relations (Greenhow &

Robelia, 2009; Ross et al., 2009; Bargh & McKenna, 2004). Secondly, they make available virtual support groups and communities for social purposes, which make it possible for users to establish new connections around their shared interests or situations (Greenhow & Robelia, 2009; Kamel Boulos & Wheeler, 2007). Thirdly, the ability of virtual environments to facilitate emotional support, and the possibility for individuals to feel comfortable with having deeper personal conversations about their problems, beyond face-to-face limitations and restrictions (Hampton et al, 2012; Ross et al., 2009). Finally, the increased sense of self esteem which is enabled (Mauri et al, 2011; Burke et al, 2010; Toma, 2010). All these factors contribute towards a person's sense of wellbeing. Other aspects of Facebook which may have some impact upon wellbeing could include looking at photos and wall postings, particularly where this activity relates to positive reminiscing.

**2.1 Positive Reminiscing and Self Soothing**

Looking at meaningful photos is a traditional method used in therapies to promote improved mood. For example, Reminiscent Therapy (RT) is a popular method used in promoting positive mood and well being, and reduces the sense of feeling alone for people with dementia. It involves using meaningful prompts, including photos, music and recordings, as an aid to remembering life events (Norris, 1986). Some research states that it has been useful in reducing depression (Scogin F *&* McElreath*,* 1994*)* as well as being an important tool to facilitate socialization.

Whilst it has been predominantly utilized in people with dementia, there could be scope for applying the theory of RT in other mental health conditions, particularly where depression and general low mood are common. This could potentially induce a „self soothing" process which could lend itself well to people who struggle with day to day living as a result of low mood, or indeed who experience the occasional „off-day". The act of „self soothing", that is calming us down, is in fact one of the hardest things to do when you have mental health problems. Yet the capability to be able to calm oneself down, to essentially self soothe, would be advantageous to people with mental health problems, and could potentially prevent problems from escalating, if only by means of a distraction.

## 3  Method

The research presented here is specifically interested in whether activities that relate to reminiscing, impact upon emotional wellbeing. A study was designed to look at people's activities on Facebook and whether these specific interactions impact upon wellbeing. In addition, it seeks to understand whether there is any indication that looking back at photos and wall posts is beneficial to people with mental health problems. This is based upon the theory that positive reminiscing has been shown to promote a sense of positive wellbeing in people with dementia (Norris, 1986; Scogin F *&* McElreath*,* 1994*)*. The research then presents the following two hypotheses:

**H1**  Looking at „wall posts" and photos shared on Facebook improves mood.

**H2** People who have experienced mental health problems will experience a greater „self soothing" effect, from looking back at wall posts and photos, than those who have not experienced mental health problems.

Facebook users were invited to complete an online survey, hosted by Survey Monkey (www.surveymonkey.net), an online survey hosting site, and was conducted in December 2012. The study was facilitated by Facebook. The researchers used their personal and University profile pages to post the link to the survey on their Facebook pages, which also included a brief explanation of the purpose of the study. The target demographic included young people of University age, as well as friends and friends of friends of the researchers. A total of 144 attempted the survey, with 135 fully completing it.

## 4. Results

The research instrument gathered information on the following areas:

1. Demographics and other descriptive data that included: frequency of accessing Facebook and whether participants owned a smart phone, as well as history of mental health problems.
2. Activities on Facebook that make people feel better, including frequency of usage.

**4.1 Demographics and General Facebook Usage**

Demographic data was collected on gender and ownership of smart phones. The survey sought to identify whether participants had experienced any degree of mental health problems. This item is particularly relevant, given that the research endeavors to compare results of the study between those that have, and those that have not, experienced mental health problems. This is in relation to whether there is any indication that people, who have experienced some degree of mental illness, derive a self soothing effect from looking at photos and wall posts. 39% of the participants stated that they had experienced mental health problems previously. The data can be seen in Table 1 below.

**Table 1**. Demographics (N=134)

|  | Mean or % (n) |  |
|---|---|---|
| Age | 34 |  |
| Gender |  |  |
|     Male | 54.5% | (73) |
|     Female | 45.5% | (61) |
| Experienced mental health issues | 39% | (53) |
| Owns a smart phone | 80% | (110) |
| Always carries phone with them | 94% | (106) |

Date was also collected and measured on the frequency of Facebook usage and behavior of participants, in relation to accessing Facebook via their phones, as shown in Table 2 below. Individual items were ranked using Likert scale, ranging from 1=strongly disagree to 5=strongly agree. The scales are presented by taking the mean value of items where the lowest possible value equals 1 and the highest possible value equals 5.

The results show that 86% of participants access Facebook more than once a day. The mean values show a tendency towards regularly accessing Facebook via phones (mean value = 3.51), with a significant majority ensuring that they always carry their phones with them (mean value = 4.63). Participants tend to also use Facebook as a distraction tool (mean value = 3.09).

**Table 2**. Summary Statistics for Facebook Use.

| Items | Mean or % |
|---|---|
| Uses Facebook more than once a day | 86% |
| I regularly access Facebook on my phone | 3.51 |
| I like being able to stay in touch with Facebook on my phone | 3.28 |
| I do not like to use Facebook on my phone | 2.45 |
| I prefer to use Facebook on my phone (than via desktop PC/iPad/other) | 2.42 |
| I always carry my phone with me | 4.63 |
| I use Facebook as a distraction tool | 3.09 |

**4.2 Facebook Activities and their Impact Upon Wellbeing**

In this section, we explore the following hypothesis:

**H1** Looking at „wall posts" and photos shared on Facebook improves mood.

Data was collected on participants' behavior on Facebook and the extent to which they participated in each activity. Individual items were ranked using a 5 point scale. 1= not at all; 2= a few times; 3= sometimes; 4= frequently and 5= almost always. The scales are presented by taking the mean value of items where the lowest possible value equals 1 (not at all) and the highest possible value equals 5 (almost always). The percentage of participants that have indicated at least „sometimes" for each activity is also shown in Table 3 below.

**Table 3**. Summary Statistics for Frequency of Facebook Activities

| Items | Mean | % |
|---|---|---|
| Looking back on wall posts | 3.76 | 86 |
| Playing games | 1.54 | 16 |
| Updating your status | 2.56 | 50 |
| Looking back at photos posted on your wall | 3.24 | 75 |
| Using Facebook messenger | 2.86 | 56 |
| Looking back at photos you have previously posted | 2.62 | 54 |

The results suggest that reading wall posts as the activity most frequently performed (mean value = 3.76) and playing games as being the least frequently performed (mean value = 1.54). 86% of participants stated that they read wall posts frequently, compared to 16% of participants who played games regularly. Whilst looking at photos isn"t an activity carried out as frequently as reading wall posts, the data does indicate that this is an activity carried out more frequently than playing games; updating status and using messenger.

Data was then gathered on which Facebook activities helped participants in improving their mood, when feeling low. Results can be seen in table 4 below. Participants were finally asked about the ease of accessing favourite photos and posts, which can be seen in table 5. Individual items were ranked using Likert scale, ranging from 1=strongly disagree to 5=strongly agree. The scales are presented by taking the mean value of items where the lowest possible value equals 1 (strongly disagree) and the highest possible value equals 5 (strongly agree).

**Table 4.** Summary Statistics of Facebook Activities that improve mood

| Items | Mean | % |
|---|---|---|
| Looking back on wall posts | 3.07 | 76 |
| Playing games | 2.04 | 32 |
| Updating your status | 2.73 | 58 |
| Looking back at photos posted on your wall | 3.14 | 73 |
| Using Facebook messenger | 2.91 | 64 |
| Looking back at photos you have previously posted | 3.13 | 71 |

The results suggest that the three highest activities that improve mood when feeling low are: looking back on wall posts (mean value = 3.07; looking back on photos previously posted (3.13) and looking at photos others have posted (mean value = 3.14). Looking back on wall posts is shown to be the most significant activity in improving mood. Results of the data also show that a significant number of participants are not able to locate favourite wall posts and photos, yet would like to be able to do so.

**Table 5**. Summary Statistics ease in accessing favourite photos and posts

| Items | Mean | % |
|---|---|---|
| It is not easy to locate my favourite photos I have posted | 3.07 | 75 |
| I would like to be able to access these photos | 3.42 | 88 |
| It is not easy to locate favourite photos others have posted | 3.35 | 84 |
| I would like to be able to access these photos. | 3.45 | 86 |
| It is not easy to locate my favourite wall comments | 3.43 | 85 |
| I would like to be able to access these wall comments | 3.47 | 86 |

These results correlate with the previous results, shown in table 4, where the data indicates that looking at photos and wall posts, does impact positively upon wellbeing.

### 4.3 The self soothing effect of Facebook activities on people with mental health issues

In this section, we explore the following hypothesis:

**H2:** People who have experienced mental health problems will experience a greater self soothing effect from looking back at wall posts and photos, than those who have not experienced mental health problems.

In table 4 above, mean values related to the self soothing effect of each Facebook activity were presented. The results showed that looking back at photos and wall posts promoted increased self soothing, when feeling low in mood. These were compared to other activities, including updating status, playing games and using Face book messenger. Further to this, the research sought to evaluate whether there was any significant difference in the effect of self soothing facilitated from looking back at photos and wall posts on Facebook, specifically between people who have experienced mental health problems and those that have not. As already stated previously, this is based upon the theory that positive reminiscing has been shown to promote a sense of positive wellbeing in people with dementia (Scogin F & McElreath, 1994). A positive significant difference with the group of participants indicating a history of mental health problems would lend further validity to the the theory.

A two sample t-Test, for difference of the population means (equal variances) was applied, to see whether there was any significant difference between the two groups; in relation to the self soothing effect that looking back upon wall posts and photos has, when feeling low in mood. Group A represents participants who have never experienced mental health problems. Group B included participants who have indicated that they have previously, or currently, experience mental health problems.

The null hypothesis predicted that there will be no significant difference between the mean values of the two groups. As in the previous results, individual items were ranked using Likert scale, ranging from 1=strongly disagree to 5=strongly agree.

**Table 6.** Looking back at wall posts; comparison of self soothing effect

| **Activity: Looking back at wall posts** | *Group A* | *Group B* |
|---|---|---|
| Mean | 2.90625 | 3.29787234 |
| Variance | 0.943452381 | 0.909343201 |
| P(T<=t) one-tail | 0.01835153 | |
| t Critical one-tail | 1.658953459 | |

The scales are presented by taking the mean value of items where the lowest possible value equals 1 (strongly disagree) and the highest possible value equals 5 (strongly agree).

**Table 7.** Looking at photos others have posted; comparison of self soothing effect

| Activity: Looking back at photos others have posted | Group A | Group B |
|---|---|---|
| Mean | 2.90625 | 3.446808511 |
| Variance | 0.975198413 | 0.948196115 |
| P(T<=t) one-tail | 0.002486147 | |
| t Critical one-tail | 1.660234327 | |

**Table 8.** Looking back at photos I have posted; comparison of self soothing effect

| Activity: Looking back at photos I have posted | Group A | Group B |
|---|---|---|
| Mean | 2.96875 | 3.333333333 |
| Variance | 1.014880952 | 1.24822695 |
| P(T<=t) one-tail | 0.038979616 | |
| t Critical one-tail | 1.661051818 | |

In all three activities, the P value is less than 0.05; therefore the null hypothesis can be rejected. These results indicate that looking back on wall posts and photos, previously posted or which others have posted has more of a self soothing effect upon those that have experienced mental health problems, than those that have never experienced mental health problems. This then further contributes to the validity of reminiscent type activities facilitating self soothing for people who are experiencing low mood.

## 5. Discussions

In considering the original research questions, the results of the study do indicate that activities involving reminiscing have a positive impact upon wellbeing. Moreover, looking back on photos and wall posts was seen to provide a greater self soothing effect, when participants were feeling low in mood, than other Facebook activities. This is further supported in that a significant number of participants were not able to easily access „favorite" wall posts and photos, and yet would like to be able to do so. In addition, the activity of looking back on photos and wall posts was carried out more frequently by participants than other activities, such as playing games, updating status and using messenger. This suggests that the activity of looking back upon photos and wall posts is a popular activity, as well as having a positive impact upon emotional wellbeing.

To further explore the theory of which Facebook activities were deemed to promote self soothing, we examined whether participants who have experienced mental health problems were able to derive a greater sense of self soothing from looking back on photos and wall posts, than those participants who have never experienced mental health problems. For each of these reminiscent type activities, the statistically analyzed results suggest a positive indication towards the potential of self soothing derived.

In spite of the positive results, there are shortcomings in this research, given the limited number of participants. Whilst statistical analysis was enabled with the number involved, the results can really only be viewed as indicative. Further research is necessary to evaluate the self soothing effect of reminiscent type activities.

The research presented here, is part of a larger study, exploring the potential of mobile applications that facilitate self soothing for different user groups (Good et al, 2012), particularly by incorporating reminiscent type (RT) therapy. Based upon RT, the application would contain meaningful memorabilia including photographs and notes. Essentially, this memorabilia would be „favorite" items that could be easily accessible and potentially promote positive mood. The significance of this research is therefore in the development of an application which can facilitate self soothing and subsequently improve a sense of wellbeing.

## References

1. Burke, M., Marlow, C., Lento, T. (2010). Social Network Activity and Social Well-Being. ACM CHI 2010: Conference on Human Factors in Computing Systems, 1909-1912
2. Cheung, C.M.K., Lee, M.K.O. (2010). A Theoretical Model of Intentional Social Action in Online Social Networks. *Decision Support Systems, 49*(1), 24-30
3. Good, A., Wilson, C., Ancient, C., Sambhanthan, A. (2012).A Proposal To Support Wellbeing in People With Borderline Personality Disorder: Applying Reminiscent Theory in a Mobile App. The ACM conference on Designing Interactive Systems
4. Greenhow, C., Robelia, B. (2009). Old Communication, New Literacies: Social Network Sites as Social Learning Resources. *Journal of Computer Mediated Communication, 14*, 1130-1161.
5. Hampton, K., Goulet, L., Marlow, C., Rainee, L. (2012). Why Most Facebook users get more than they give. Pew Internet.
6. Kamel Boulos, M. N., Wheeler, S. (2007). The emerging Web 2.0 social software: an enabling suite of sociable technologies in health and healthcare education. *Health Information and Libraries Journal, 24*, 2–23.
7. Mauri, M., Cipresso, P., Balgera, A., Villamira, M. & Riva, G. (2011). Why Is Facebook So Successful? Psychophysiological Measures Describe a Core Flow State While Using Facebook. *Cyberpsychology, Behavior & Social Networking.* 14(12)
8. Neal*, D.M., & McKenzie, P*.J. (2010) "I Did Not Realize So Many Options Are Available": Cognitive Authority, Emerging Adults, and e- Mental Health, *Library & Information Science Research* (2010)
9. Norris, A.D. (1986). *Reminiscence* with Elderly People. London*: Winslow,*
10. Ross, C., Orr, E.S., Sisic M., Arseneault, J.M., Simmering, M.G. and Orr, R.R. (2009) Personality and Motivations Associated With Facebook Use. *Computers in Human Behavior, 25*(2), 578–586
11. Scogin F*,* McElreath L. Efficacy of Psychosocial Treatments for Geriatric Depression: A

Quantitative Review. Journal of Consulting & Clinical Psychology *(*1994) 62*:*69-74
12. Taylor, C.B., Luce, K.H. (2003). Computer- and Internet-Based Psychotherapy interventions. *Current Directions in Psychological Science*, 12,18–22
13. Toma, C., (2010). Affirming the Self Through Online Profiles: Beneficial Effects of Social Network Sites. *Proceedings of the SIGCHI Conference on Human Factors in Computing System.* ACM, New York
14. Van De Belt, T. H., Engeleni, L., Berbent, S. A. A., & Schoonhoven, L. (2010). Definition of Health 2.0 and Medicine 2.0: A Systematic Review. *Journal of Medical Internet Research, 12*(2)
15. Van Uden-Kraan, C.F., Drossaert, C.H.C., Taal, E., Lebrun, C.E.I., Drossaers-Bakker, K.W., Smit, W.M., Seydel, E.R. and Van de Laar, M.A.F.J. (2008). Coping With Somatic Illnesses in Online Support Groups. Do the Feared Disadvantages Actually Occur? *Comput Human Behav, 24*, 309–324.